\documentclass[prb,twocolumn,superscriptaddress,showpacs,preprintnumbers,amsmath,amssymb]{revtex4}

\usepackage{graphicx}
\usepackage{dcolumn}
\usepackage{bm}
\usepackage{enumerate}

\begin{document}

\bibliographystyle{apsrev} 

\title{High-fidelity quantum memory utilizing inhomogeneous nuclear polarization \\in a quantum dot}

\author{Wenkui Ding}
\affiliation{School of Physics and Technology, Wuhan University, Wuhan, Hubei 430072, China}
\author{Anqi Shi}
\affiliation{School of Physics and Technology, Wuhan University, Wuhan, Hubei 430072, China}
\author{J. Q. You}
\affiliation{Beijing Computational Science Research Center, Beijing 100084, China}
\affiliation{Synergetic Innovation Center of Quantum Information and Quantum Physics, University of Science and Technology of China, Hefei, Anhui 230026, China}
\author{Wenxian Zhang}
\email[Corresponding email: ]{wxzhang@whu.edu.cn}
\affiliation{School of Physics and Technology, Wuhan University, Wuhan, Hubei 430072, China}
\affiliation{Kavli Institute for Theoretical Physics China, CAS, Beijing 100190, China}

\date{\today}

\begin{abstract}
We numerically investigate the encoding and retrieval processes for a quantum memory realized in a semiconductor quantum dot, by focusing on the effect of inhomogeneously polarized nuclear spins whose polarization depends on the local hyperfine coupling strength. We find that the performance of the quantum memory is significantly improved by the inhomogeneous nuclear polarization, as compared to the homogeneous one. Moreover, the narrower the nuclear polarization distribution is, the better the performance of the quantum memory is. We ascribe the performance improvement to the full harnessing of the highly polarized and strongly coupled nuclear spins, by carefully studying the entropy change of individual nuclear spins during encoding process. Our results shed new light on the implementation of a quantum memory in a quantum dot.
\end{abstract}

\maketitle

\section{Introduction}

A key ingredient of quantum computation and quantum communication is quantum memory, which may be implemented in many physical systems, such as cold atomic gases~\cite{duan2001long, Kuzmich}, nuclear spin systems~\cite{loss1998quantum}, semiconductor quantum dots (QD)~\cite{Taylor03,RevModPhys.79.1217,PhysRevLett.103.010502, PhysRevB.76.045218, wang2012self}, and so on. Among these systems, the QD-based quantum memory, which uses both electron spin and nuclear spins, exhibits potential advantages: Long storage time, fast encoding and retrieval of the stored quantum state, and ready to scale-up with current semiconductor fabrication techniques~\cite{Merkulov02,kane1998silicon,loss1998quantum,Taylor03,Dobrovitski06}.

The quantum memory protocol proposed for a QD utilizes the easy controllability of the electron spin and the long coherence time of the nuclear spins~\cite{Taylor03,Salis2001, elzerman2004single}. For the perfectly polarized nuclear spins, the fidelity of a quantum state after encoding, storage, and retrieval is approaching 100\%. While for partially polarized nuclear spins, the fidelity reduces linearly with the decrease of the average nuclear polarization~\cite{Dobrovitski06,Khaetskii2002,Prokofev2000}. In order to achieve a reasonable fidelity, for example 80\%, the average nuclear polarization is required to be above 80\%, which is beyond the availability of current QD experiments whose record is 68\%  with optical pumping methods~\cite{gammon2001electron,dnp2009,Asshoff11,PhysRevB.15.5780,imamoglu2003}.

To alleviate the high nuclear polarization requirement while keeping the reasonable fidelity, it is a possible way to employ the inhomogeneous nuclear polarization, which could be prepared through dynamic nuclear polarization~\cite{Reilly08, Dixon1997, PhysRevLett.100.067601, Lukin10, Zhang10, Zhang13arxiv}. On one hand, the nuclear polarization after dynamic nuclear polarization is proportional to the square of the local hyperfine coupling constant at short times and is saturated at long times, so the polarizations of the strongly coupled nuclear spins are much higher than those of the weakly coupled ones. On the other hand, the strongly coupled nuclear spins play a more important role in encoding and retrieval process than the weakly coupled nuclear spins. The effective polarization for a quantum memory must be weighted by the local hyperfine coupling constants in a certain way. In contrast to a QD with homogeneously polarized nuclear spins, the effective polarization is higher for a QD with inhomogeneously polarized nuclear spins. Thus, the fidelity of a quantum memory with  inhomogeneous polarization may be higher than that of a quantum memory with homogeneous polarization.

In this paper, we investigate systematically with numerical method the performance of a QD-based quantum memory with homogeneous or inhomogeneous polarization. The minimal fidelity of an arbitrary quantum state after the encoding and retrieval process is compared quantitatively for the homogeneous and inhomogeneous polarizations. To understand the superiority of the inhomogeneously polarized QD, we further numerically investigate the von Neumann entropy change of each nuclear spin from the viewpoint of quantum information theory~\cite{nielsen2010quantum}.

The paper is organized as follows.
In Sec.~\ref{sec:qm}, we briefly review the quantum memory protocol for a QD. Detailed comparisons of numerical results for homogeneous and inhomogeneous polarizations are presented in Sec.~\ref{sec:nr}. In Sec.~\ref{sec:iepy}, we discuss the entropy change of nuclear spins during the encoding stage, in order to understand the superior performance of a QD with inhomogeneous polarization. Finally, the conclusion is drawn in Sec.~\ref{sec:con}.

\section{Quantum memory protocol for a QD}
\label{sec:qm}

\begin{figure}
\includegraphics[width=3.25in]{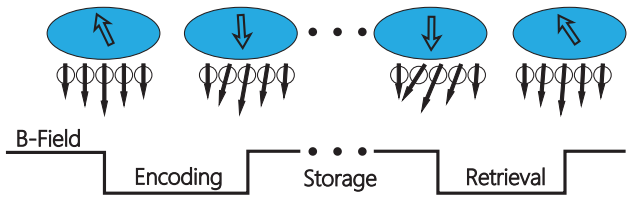}
\caption{\label{fig:qmp} (Color online) Three stages of a quantum memory protocol for an {\it inhomogeneously} polarized QD. (I) Encoding: mapping a quantum state of the electron spin onto a collective state of nuclear spins. (II) Storage: ejecting the electron from the QD and the nuclear spin state is preserved for a required period. (III) Retrieval: injecting another electron spin in $|\downarrow\rangle$ state and mapping back the initial quantum state. The external magnetic field is tuned on resonance with the Overhauser field during the encoding and retrieval stages. Strongly coupled nuclear spins play more important roles in this quantum memory protocol.}
\end{figure}

The goal of a quantum memory is to first encode a quantum state into a well-isolated system, whose coherence time is long~\cite{duan2001long,opinhomo12,PhysRevA.84.063810}, for instance, cold atomic gases~\cite{Kuzmich} or nuclear spins~\cite{Taylor03}. After a desired storage time, the state is then retrieved with a high fidelity. For a QD, the combination of the highly controllable electron spin and the long coherence-time nuclear spins makes it an ideal candidate for a quantum memory.

In a QD, a complete quantum memory cycle consists of three steps, as shown in Fig.~\ref{fig:qmp}. The first step is to encode the information carried by the electron spin, i.e., an arbitrary initial state $\alpha |\uparrow\rangle + \beta |\downarrow\rangle$, into a collective state of nuclear spins in the QD. The dynamics of the electron and $N$ nuclear spins during the encoding (and the retrieval) stage is governed by the following Hamiltonian,
\begin{equation}\label{eq:ham}
H = g_e^* \mu_B B_0 S^z + \sum_{k=1}^N A_k {\mathbf I}_k \cdot {\mathbf S}
\end{equation}
where the first term corresponds to the Zeeman energy of the electron spin $\mathbf S$ in an external magnetic field $B_0$ along $z$-axis, with $g_e^*$ being the g-factor of the electron and $\mu_B$ the Bohr magneton. The second term corresponds to the Fermi contact hyperfine interaction where $A_k = A v_0 |\psi({\mathbf r}_k)|^2$ ($k=1,2, \cdots, N$) is the coupling strength between the electron spin and the $k$th nuclear spin, with $A$ being  the one-electron hyperfine interaction constant, $v_0$ the volume of a unit cell, and $|\psi({\mathbf r}_k)|^2$ the electron density profile at site ${\mathbf r}_k$ which usually varies in a Gaussian form in a QD under typical experimental conditions~\cite{johnson2005triplet, Koppens2005, Petta2005}.

To efficiently encode the electron spin state into the nuclear spins, the magnetic field $B_0$ is tuned on resonance with the Overhauser field $B_{\rm{over}} = \sum_k A_k \langle I_k^z \rangle$. As shown for perfectly polarized nuclear spins~\cite{Taylor03}, the effective Hamiltonian becomes
\begin{equation}
\label{eq:sham}
H = \sum_{k=1}^N {(A_k/2)({I}_k^+ {S}^- + {I}_k^-{S}^+)}
\end{equation}
which dominates spin exchange between the electron and nuclear spins. The raising and lowering operators are defined as $S^{\pm}=S^x{\pm}iS^y$, and $I_k^{\pm}=I_k^x{\pm}iI_k^y$. After half a period of Rabi oscillations between two basis states $|\uparrow\rangle_e \otimes |0\rangle_n$ and $|\downarrow\rangle_e \otimes |1\rangle_n$, where $|0\rangle_n \equiv |\downarrow\downarrow\cdots\downarrow\rangle_n$ and $|1\rangle_n\equiv (1/\Omega) \sum _{k=1}^N A_k |\downarrow \cdots \uparrow_k \cdots \downarrow \rangle_n$ with $\Omega = (\sum_{k=1}^N |A_k|^2)^{1/2}$ being the Rabi angular frequency, the state becomes $|\downarrow\rangle_e\otimes(\alpha|1\rangle_n+i\beta|0\rangle_n)$ and the initial electron spin state is mapped onto the final collective nuclear spin state.

Once the electron spin state is mapped onto the nuclear spins, the electron is ejected from the QD. The collective nuclear spin state is preserved for a desired period~\cite{Taylor03,Dobrovitski06}. The following retrieval process is in fact an inverse of the encoding process, i.e., another electron spin in $|\downarrow\rangle$ state is injected into the QD and the whole system evolves under the same Hamiltonian Eq.~(\ref{eq:ham}) for half a period of the Rabi oscillations. The initial electron spin state is then restored after the complete quantum memory protocol.

In a real QD, the required perfect nuclear polarization is impossible to realize. To estimate the effect of imperfect nuclear polarization, studies have been done both analytically with an assumption of uniform hyperfine coupling and numerically with nonuniform hyperfine coupling~\cite{Taylor03,Dobrovitski06}. These results indicate that the minimal fidelity of the quantum memory protocol in the worst case decays linearly with the decrease of the polarization. Noticeably, the minimal fidelity drops below 80\% even at a pretty high nuclear polarization of 80\%. To keep the minimal fidelity but reduce the nuclear polarization to an experimentally accessible value (68\%), further efforts are needed.

We notice that the hyperfine coupling is nonuniform in a QD and the nuclear polarization is not necessarily uniform after dynamic nuclear polarization (the strongly coupled nuclear spins actually acquire higher polarization)~\cite{Zhang10, Lukin10, Zhang13arxiv, Lukin13, Yao_EPL_2010}. To fully utilize the dynamically polarized nuclear spins, we next investigate the performance of the quantum memory protocol with an inhomogeneous polarization, hopefully to reduce the average polarization while keeping the same minimal fidelity.

\section{Effect of inhomogeneous polarizations}
\label{sec:nr}

\begin{figure}
\includegraphics[width=3in]{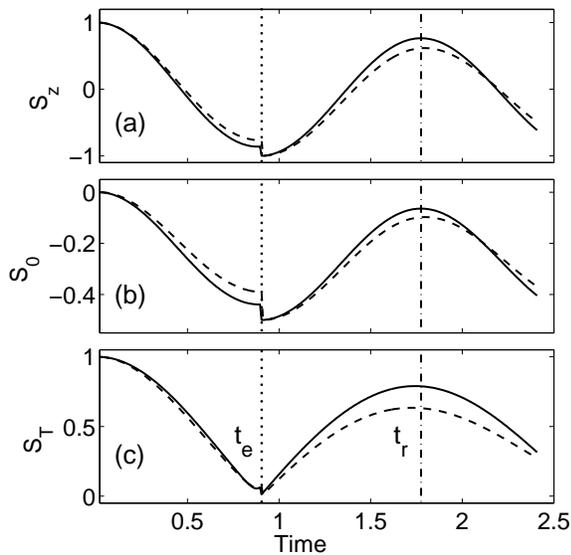}
\caption{\label{fig:sd} Typical evolutions of (a) $s_z$, (b) $s_0$, and (c) $s_T$ for $\Delta P$=0.2 and $N$=20. The solid and dashed lines denote the results for the inhomogeneous and homogeneous polarization cases, respectively. The vertical dotted and dash-dotted lines denote respectively the ejection time $t_e$ and the retrieval time $t_r$. After the whole process, the retrieved state is closer to the initial electron spin state in the inhomogeneous case. }
\end{figure}

\begin{figure}
\includegraphics[width=3in]{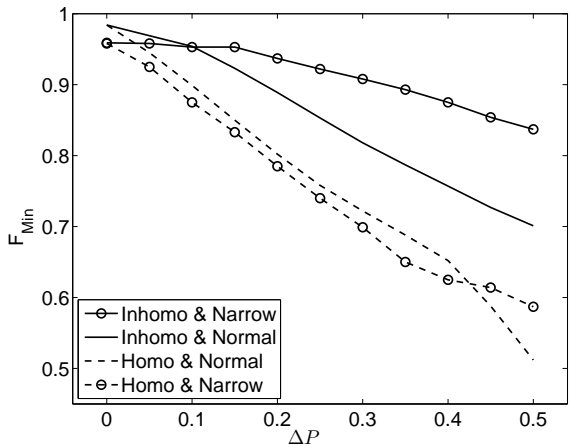}
\caption{\label{fig:fm} Minimal fidelity as a function of $\Delta P$ for four cases: (I) Inhomogeneous polarization with a narrow width of $A_k$ (solid line with circles); (II) Inhomogeneous polarization with a normal width of $A_k$ (solid line); (III) Homogeneous polarization with a normal width of $A_k$ (dashed line); (IV) Homogeneous polarization with a narrow width of $A_k$ (dashed line with circles). Obviously, a quantum memory with inhomogeneous polarization exhibits better performance.}
\end{figure}

For an inhomogeneously polarized QD with nonuniform hyperfine couplings, it is challenging to obtain analytical solution to the encoding and retrieval dynamics~\cite{PhysRevB.70.195340,exsolution07,gaudin1976diagonalization,Faribault13}. We thus employ numerical method to simulate the encoding and retrieval stages of the quantum memory protocol.

The initial nuclear spins are prepared in an inhomogeneously polarized state with the $k$th spin's polarization as $p_k = \tanh(\beta A_k^2)$, where $\beta$ is an adjustable parameter. Theoretical predictions show that such an initial state may be experimentally realized by employing the dynamic nuclear polarization method under the condition of short enough period~\cite{Zhang10,Zhang13arxiv}. We numerically obtain this state by acting the operator $\exp(\beta \sum_k A_k^2 I_k^z)$ on an initial random state $|r\rangle$ of the nuclear spins~\cite{Dobrovitski06}, where $|r\rangle = \sum_{i=1}^{2^N} c_i |i\rangle$ is a linear combination of basis states of all nuclear spins and $c_i$ are independent identically distributed random complex numbers obeying $\sum_{i=1}^{2^N} |c_i|^2 = 1$. Such a superposition is an exponentially accurate representation of the maximally mixed state and in our simulations creates errors of about $0.1\%$. To find the minimal fidelity $F_{\rm{min}}=\min_{\psi_0}[\langle \psi_0 | \rho(t_r) |\psi_0\rangle]$ with $|\psi_0\rangle$ being the initial electron state and $\rho(t_r)$ the final mixed electron state, two states initially along $z$ and $x$ axes, respectively, are simulated.

At the beginning of the encoding stage, the external magnetic field is tuned numerically to reach the largest value of $F_{\rm{min}}$ for a given nuclear polarization (more details are presented in the Appendix~\ref{sec:mf}). This magnetic field is fixed during the later encoding and retrieval stages. The value of this optimized magnetic field is close to the Overhauser field $B_{\rm{over}}$. For the small system size that we consider here, $N=20$, the external magnetic field is slightly larger than $B_{\rm{over}}$ and depends weakly on the nuclear polarization, due to the finite size effect.

In order to clearly illustrate the advantages of the inhomogeneous polarization, we adopt the same $A_k$'s as in Ref.~\onlinecite{Dobrovitski06}, where $N=4\times 5$ nuclear spins are placed in a rectangular lattice with the lattice constants $a_x$ and $a_y$. The constant $A_k$ is in a two-dimension Gaussian form with the widths $w_x$ and $w_y$ and a shifted centre, $A_k \propto \exp[-(x-x_0)^2/w_x^2 - (y-y_0)^2/w_y^2]$ with $x_0 = 0.1 a_x$ and $y_0 = 0.2 a_y$. Two widths of the Gaussian form are employed. For a normal width, $w_x/a_x = 3/2$ and $w_y/a_y = 2$ along $x$ and $y$ axes, respectively, with the largest constant $A_k$ being 0.96. For a narrow width, $w_x/a_x = 3/(2\sqrt{2})$ and $w_y/a_y = \sqrt{2}$, with the largest $A_k$ being 0.92.

We employ the method of the Chebyshev polynomial expansion of the evolution operator to evolve the coupled many-spin system~\cite{Dobrovitski03}. With this method, we may simulate the dynamics of up to 30 spins. As discussed in Ref.~\onlinecite{Dobrovitski06}, the results for 20 nuclear spins are almost identical to the results for $10^4$ nuclear spins within a Rabi oscillation, so $N=20$ is reliable to simulate the realistic QD cases. Therefore, we also consider $N=20$ nuclear spins in this paper. To extract the minimal fidelity, we need to monitor the following three observables, $s_{x,y,z} = \rm{tr}\{\hat S^{x,y,z} \rho(t)\}$, where $\rho(t)$ is the density matrix of the coupled system at time $t$. We also define the transverse and longitudinal components of the electron spin $s_T = \sqrt{s_x^2 + s_y^2}$ and $s_0=s_z$, for the specific initial electron state along the $x$ direction.

Typical evolution of the electron spin in the encoding and retrieval stages is illustrated in Fig.~\ref{fig:sd}. The ejection time $t_e$ corresponds to the minimal $s_z$ during the encoding stage. The nonunitary ejection of the electron is calculated numerically as a von Neumann projection and the left nuclear spin state is
$\rho_n(t_e) = tr_e[\rho(t_e)]=\langle\uparrow|\,\rho\,|\uparrow\rangle+\langle\downarrow|\,\rho\,|\downarrow\rangle$, which traces out the electron's degree of freedom. Right after the injection of the second electron in $|\downarrow \rangle$ state at the beginning of the retrieval stage, the density matrix of the coupled system becomes $|\downarrow\rangle \langle \downarrow| \otimes \rho_n(t_e)$. The final retrieval time $t_r$ is located at the largest value of $F_{\rm{min}}$. By employing the idea of quantum process tomography and using the decomposition of an arbitrary $2\times 2$ matrix into the Pauli matrices and the identity matrix, as proved in Ref.~\onlinecite{Dobrovitski06}, the minimal fidelity $F_{\rm{min}}$ is straightforwardly calculated by finding the minimum value of the following three quantities,
\begin{equation}
\begin{aligned}
&f_1=\frac{1+s_z}{2},\\
&f_2=\frac{1+s_z-2s_0}{2},\\
&f_3=\frac{1}{2}\left[1+s_T-{s_0^2\over 4(s_z-s_0-s_T)}\right].
\end{aligned}
\nonumber
\end{equation}

Figure~\ref{fig:sd}(a) shows the longitudinal component of the electron spin $s_z$ for $\Delta P = 0.2$ with inhomogeneous and homogeneous polarizations, where $\Delta P = 1- (1/N)\sum_{k=1}^N p_k$ is the deviation from the perfect polarization. The initial electron spin state is along the $z$ axis. Comparing the value of $s_z$ at time $t_r$ to the initial value for the two cases, we find that the inhomogeneous one is closer. Similarly, for an initial state along the $x$ axis, as shown in Fig.~\ref{fig:sd}(b) and (c), $s_0$ and $s_T$ both show closer value at time $t_r$ to their initial value in the inhomogeneous case. We also note that the oscillation amplitude in the inhomogeneous case is larger than that in the homogeneous one, indicating that more information is encoded into the nuclear spins in the inhomogeneous case.

The minimal fidelity $F_{\rm{min}}$ calculated with $s_z$, $s_0$, and $s_T$ at time $t_r$ are presented in Fig.~\ref{fig:fm} for various $\Delta P$'s in homogeneous and inhomogeneous cases. The dashed lines in Fig.~\ref{fig:fm} corresponding to the homogeneous cases are essentially the reproduction of the main results in Ref.~\onlinecite{Dobrovitski06}. As a comparison to our results for inhomogeneous polarizations in this paper (solid lines in Fig.~\ref{fig:fm}): First, $F_{\rm{min}}$'s in the inhomogeneous cases are larger than in the homogeneous cases; Second, for the inhomogeneous cases with different widths, the narrow width situation exhibits larger $F_{\rm{min}}$'s than the normal width, while for the homogenous cases $F_{\rm min}$ is almost independent of the width. Noticeably, in the case of inhomogeneous polarization with narrow width of $A_k$'s, the value of $F_{\rm{min}}$ is still above 80\% even at $\Delta P = 0.5$, which corresponds to a nuclear polarization well below the experimentally accessible value 68\%.

To explain the superior performance in the inhomogeneous case, we notice that the errors, $1-F_{\rm{min}}$, are mainly caused by the strongly coupled nuclear spins in the case of homogeneous polarization~\cite{Dobrovitski06}. While in the inhomogeneous case, these strongly coupled nuclear spins have higher polarization than the average, so the main contribution to the errors is significantly suppressed and the minimal fidelity becomes larger. Similarly, the narrower the distribution of the inhomogeneous polarization is, the higher the minimal fidelity is.

\section{Entropy change of nuclear spins during encoding stage}
\label{sec:iepy}

\begin{figure}
\includegraphics[width=3.25in]{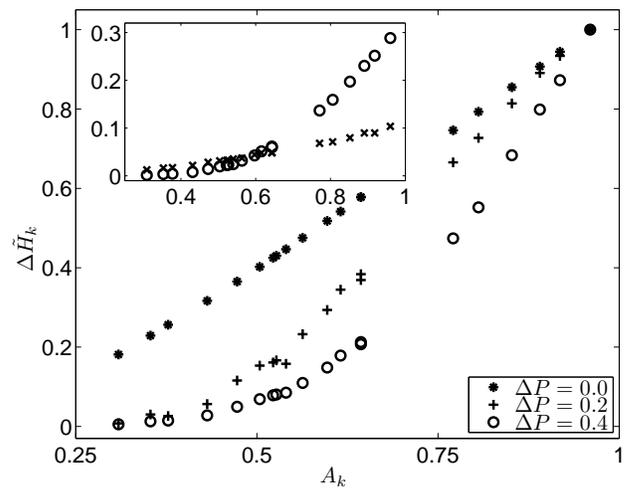}
\caption{\label{fig:epy} Dependence of $\Delta \tilde H_k$ on $A_k$ for various polarizations at optimal magnetic fields during the encoding stage. $A_k$ is in a Gaussian form with a normal width. Insert: Comparison of original $\Delta H_k$ for $\Delta P=0.4$ with homogeneous (crosses) and inhomogeneous (circles) polarizations. Strongly coupled and highly polarized nuclear spins acquire more information during the encoding stage. }
\end{figure}

\begin{figure}
\includegraphics[width=3.25in]{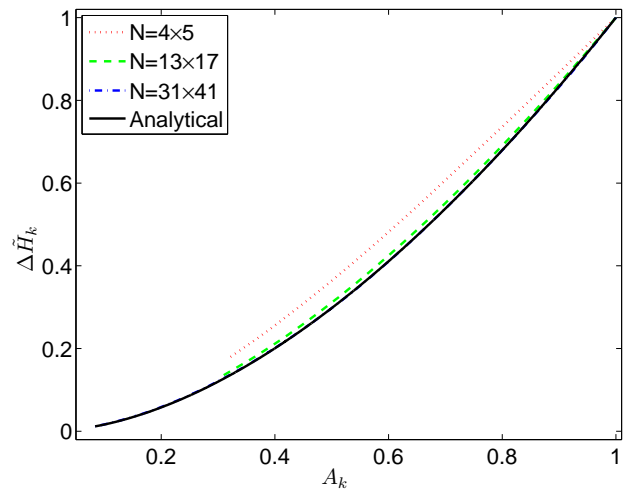}
\caption{\label{fig:epya} (Color online) Dependence of $\Delta \tilde H_k$ on $A_k$ for the perfect polarization. $A_k$'s are normalized to their largest value. The solid line, which is indistinguishable from the dash-dotted line for $N=31\times 41$, stands for the analytical limiting solution at $N\rightarrow \infty$. }
\end{figure}

During the encoding stage in the case of inhomogeneous polarization, as shown in Fig.~\ref{fig:sd}, the larger amplitude of the Rabi oscillation of the electron spin indicates that more information is written into the nuclear spins, which inspires us to further quantify the information acquired by the nuclear spins.

To measure the information acquired by each nuclear spin, we employ the change of the von Neumann entropy,
\begin{equation}
\Delta H_k=H_k(t_e)-H_k(0)
\end{equation}
where $H_k(t)=-\rm{tr}[\rho_k(t) \ln \rho_k(t)]$, with $\rho_k(t)$ being the reduced density matrix of the $k$th nuclear spin at time $t$. To compare the $\Delta H_k$'s behaviors for different polarizations, we normalize them to their largest value, i.e., $\Delta \tilde H_k = \Delta H_k / \max(\Delta H_k)$ for each nuclear polarization $\Delta P$.

We plot in Fig.~\ref{fig:epy} the normalized change of the von Neumann entropy $\Delta \tilde H_k$ for each nuclear spin as a function of the coupling strength $A_k$ for various inhomogeneous nuclear polarizations with the initial electron spin along the $z$ axis. As shown clearly in the figure, strongly coupled (also highly polarized) nuclear spins change their state more drastically, i.e., with larger $\Delta \tilde H_k$. It reveals that these nuclear spins acquire more information from the electron during the encoding stage. This results hold for all polarizations and are more prominent for larger $\Delta P$'s. For example, the difference of $\Delta \tilde H_k$ at larger $A_k$'s and at smaller $A_k$'s for $\Delta P = 0.4$ is larger than that for $\Delta P = 0.0$ or $0.2$. By comparing $\Delta H_k$ of the homogeneous and inhomogeneous polarizations for $\Delta P = 0.4$ in the inset of Fig.~\ref{fig:epy}, we observe that the strongly coupled nuclear spins acquire much more information in the inhomogeneous case. This may explain why the quantum memory performance is better in the inhomogeneous polarization. As a thumb rule, we note here that the overall $\Delta H_k$ becomes smaller for larger $\Delta P$'s.

For the perfectly polarized nuclear spins, we may investigate the entropy change during the encoding stage for much more nuclear spins by utilizing the conservation of the $z$ component of the total spins $S^z + \sum_{k=1}^N I_k^z$. We increase the number of nuclear spins by shrinking the lattice constant to $1/4$ and $1/10$ of their original values but keep the profile of the electron density fixed. For a consistent comparison, we normalize not only $\Delta H_k$ by the maximum $\max(\Delta H_k)$ but also $A_k$ by the maximum $\max(A_k)$.

The results for the perfect polarization are presented in Fig.~\ref{fig:epya} for three QD sizes. These curves are remarkably close to each other, manifesting that the conclusions drawn from $N=20$ nuclear spins may also applicable to larger QD sizes. For $N\rightarrow \infty$, we are actually able to obtain analytical solution.

As $N\rightarrow \infty$, the resonance condition is in fact always satisfied, so the Hamiltonian is further simplified to Eq.~(\ref{eq:sham}). For an initial state $|\uparrow\rangle_e \otimes |0\rangle_n$, the state becomes $|\downarrow\rangle_e \otimes |1\rangle_n$ after a time of $\pi/\Omega$. The reduced density matrix of each nuclear spin is expressed as
\begin{equation}
\rho _k=\frac{1}{2}\left(
\begin{array}{cc}
 1+p_k  & 0 \\
 0  & 1-p_k  \\
\end{array}
\right)
\end{equation}
where $p_k = 2\langle I_k^z \rangle = 1-2a_k^2$, with $a_k = A_k / \Omega$, is the polarization of the $k$th nuclear spin. Finally, we obtain the entropy change of each nuclear spin
\begin{equation}
\Delta H_k=-\left(1-{a_k^2}\right)\ln \left(1-{a_k^2}\right)-{a_k^2}\ln \left({a_k^2}\right).
\end{equation}
The solid line in Fig.5 shows this analytical result. As shown in the figure, this analytical curve looks indistinguishable from the one for $N=31\times 41$.

\section{Conclusion}
\label{sec:con}

To conclude, we have investigated the effect of the inhomogeneous nuclear polarization on the performance of a QD-based quantum memory. Compared with the homogeneous nuclear polarization, a QD-based quantum memory has a much higher minimal fidelity with the inhomogeneous polarization. Remarkably, the minimal fidelity can reach above 80\% even at a nuclear polarization as low as 50\%. We ascribe the superior performance in inhomogeneous polarization to the suppression of the errors mainly caused by the strongly coupled nuclear spins, whose polarizations are higher than the average polarization. We further carry out the calculations of the entropy change during the encoding stage and the results show that the strongly coupled nuclear spins indeed dominate at this stage for QD sizes varying from 20 to above 2000 nuclear spins. Our results indicate a practical way to experimentally realize a QD-based quantum memory with inhomogeneous nuclear polarizations.

\begin{acknowledgments}
This work is supported by the National Basic Research Program of China (Grant No. 2013CB922003 and 2014CB921401), the National Natural Science Foundation of China under Grant No. 11275139 and 91121015, the NSAF Grant No. U1330201, and the Fundamental Research Funds for the Central Universities.
\end{acknowledgments}

\appendix*
\section{Optimal magnet field}
\label{sec:mf}

\begin{figure}
\includegraphics[width=3in]{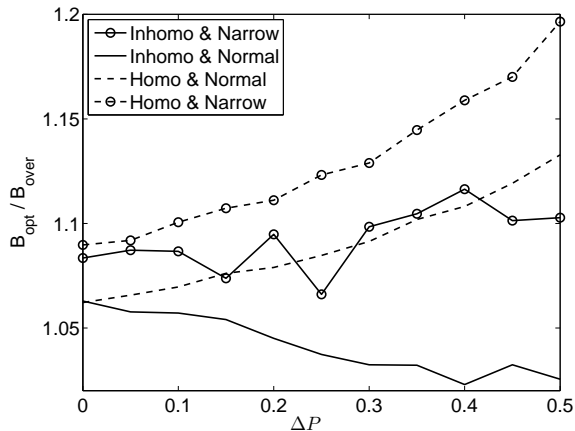}
\caption{\label{fig:opf} Numerically optimized magnetic field, normalized by the Overhauser fields, as a function of $\Delta P$ for four cases: (I) Inhomogeneous polarization with a narrow width of $A_k$ (solid line with circles); (II) Inhomogeneous polarization with a normal width of $A_k$ (solid line); (III) Homogeneous polarization with a normal width of $A_k$ (dashed line); (IV) Homogeneous polarization with a narrow width of $A_k$ (dashed line with circles). }
\end{figure}

During the evolution of the quantum memory protocol, the external magnetic field is applied to compensate the Overhauser field to fulfill the resonance condition. However, the Overhauser field is time dependent and fluctuating. Although the best external filed should also be time dependent in principle, a practical external magnetic field is constant both in our calculations and in experiments. To best meet the requirement of the resonance condition, we numerically optimize $B_0$ by searching for the largest $F_{\rm{min}}$ for both homogeneous and inhomogeneous nuclear polarizations. For a homogeneous polarization, it is analytically shown that the optimal magnetic field is~\cite{Dobrovitski06}
\begin{equation}
\label{eq:Bopt}
B_0=-B_{\rm{over}} - M_3/(2M_2g_e^*\mu_B)
\end{equation}\\
where $M_3=\sum_{k=1}^N {A_k^3}$, and $M_2=\sum_{k=1}^N {A_k^2}$. We numerically simulate the evolution and calculate $F_{\rm min}$ for 10 magnetic fields around $B_0$ with an interval of 0.1. We then fit the 10 pairs of \{$B, F_{\rm min}$\} with a quadratic function and locate the optimal magnetic field $B_{\rm opt}$ which corresponds to the largest $F_{\rm min}$. Using the optimal field $B_{\rm opt}$, we finally simulate the evolution again and obtain the results of $F_{\rm min}$ shown in Fig.~\ref{fig:fm}. For the homogeneous cases, our numerically obtained optimal magnetic fields are actually the same as analytical results~(\ref{eq:Bopt}), but for the inhomogeneous cases, we find the numerical optimal value deviates away from the analytical results when $\Delta P$ increases. We note that this deviation would become small as the number of nuclear spins increases, i.e., $B_{\rm opt} \rightarrow -B_{\rm over}$ if $N\rightarrow \infty$.

We present in Fig.~\ref{fig:opf} the optimal magnetic fields normalized by the Overhauser field $B_{\rm{over}}=\sum_{k}p_{k}A_{k}$ for various nuclear polarizations with $N=20$. These normalized optimal magnetic fields are almost constant between 1 and 1.2.



\begin{thebibliography}{39}
\expandafter\ifx\csname natexlab\endcsname\relax\def\natexlab#1{#1}\fi
\expandafter\ifx\csname bibnamefont\endcsname\relax
  \def\bibnamefont#1{#1}\fi
\expandafter\ifx\csname bibfnamefont\endcsname\relax
  \def\bibfnamefont#1{#1}\fi
\expandafter\ifx\csname citenamefont\endcsname\relax
  \def\citenamefont#1{#1}\fi
\expandafter\ifx\csname url\endcsname\relax
  \def\url#1{\texttt{#1}}\fi
\expandafter\ifx\csname urlprefix\endcsname\relax\def\urlprefix{URL }\fi
\providecommand{\bibinfo}[2]{#2}
\providecommand{\eprint}[2][]{\url{#2}}

\bibitem[{\citenamefont{Duan et~al.}(2001)\citenamefont{Duan, Lukin, Cirac, and
  Zoller}}]{duan2001long}
\bibinfo{author}{\bibfnamefont{L.-M.} \bibnamefont{Duan}},
  \bibinfo{author}{\bibfnamefont{M.}~\bibnamefont{Lukin}},
  \bibinfo{author}{\bibfnamefont{J.~I.} \bibnamefont{Cirac}}, \bibnamefont{and}
  \bibinfo{author}{\bibfnamefont{P.}~\bibnamefont{Zoller}},
  \bibinfo{journal}{Nature (London)} \textbf{\bibinfo{volume}{414}},
  \bibinfo{pages}{413} (\bibinfo{year}{2001}).

\bibitem[{\citenamefont{Chaneliere et~al.}(2005)\citenamefont{Chaneliere,
  Matsukevich, Jenkins, Lan, Kennedy, and Kuzmich}}]{Kuzmich}
\bibinfo{author}{\bibfnamefont{T.}~\bibnamefont{Chaneliere}},
  \bibinfo{author}{\bibfnamefont{D.}~\bibnamefont{Matsukevich}},
  \bibinfo{author}{\bibfnamefont{S.}~\bibnamefont{Jenkins}},
  \bibinfo{author}{\bibfnamefont{S.-Y.} \bibnamefont{Lan}},
  \bibinfo{author}{\bibfnamefont{T.}~\bibnamefont{Kennedy}}, \bibnamefont{and}
  \bibinfo{author}{\bibfnamefont{A.}~\bibnamefont{Kuzmich}},
  \bibinfo{journal}{Nature (London)} \textbf{\bibinfo{volume}{438}},
  \bibinfo{pages}{833} (\bibinfo{year}{2005}).

\bibitem[{\citenamefont{Loss and DiVincenzo}(1998)}]{loss1998quantum}
\bibinfo{author}{\bibfnamefont{D.}~\bibnamefont{Loss}} \bibnamefont{and}
  \bibinfo{author}{\bibfnamefont{D.~P.} \bibnamefont{DiVincenzo}},
  \bibinfo{journal}{Phys. Rev. A} \textbf{\bibinfo{volume}{57}},
  \bibinfo{pages}{120} (\bibinfo{year}{1998}).

\bibitem[{\citenamefont{Taylor et~al.}(2003)\citenamefont{Taylor, Marcus, and
  Lukin}}]{Taylor03}
\bibinfo{author}{\bibfnamefont{J.~M.} \bibnamefont{Taylor}},
  \bibinfo{author}{\bibfnamefont{C.~M.} \bibnamefont{Marcus}},
  \bibnamefont{and} \bibinfo{author}{\bibfnamefont{M.~D.} \bibnamefont{Lukin}},
  \bibinfo{journal}{Phys. Rev. Lett.} \textbf{\bibinfo{volume}{90}},
  \bibinfo{pages}{206803} (\bibinfo{year}{2003}).

\bibitem[{\citenamefont{Hanson et~al.}(2007)\citenamefont{Hanson, Kouwenhoven,
  Petta, Tarucha, and Vandersypen}}]{RevModPhys.79.1217}
\bibinfo{author}{\bibfnamefont{R.}~\bibnamefont{Hanson}},
  \bibinfo{author}{\bibfnamefont{L.~P.} \bibnamefont{Kouwenhoven}},
  \bibinfo{author}{\bibfnamefont{J.~R.} \bibnamefont{Petta}},
  \bibinfo{author}{\bibfnamefont{S.}~\bibnamefont{Tarucha}}, \bibnamefont{and}
  \bibinfo{author}{\bibfnamefont{L.~M.~K.} \bibnamefont{Vandersypen}},
  \bibinfo{journal}{Rev. Mod. Phys.} \textbf{\bibinfo{volume}{79}},
  \bibinfo{pages}{1217} (\bibinfo{year}{2007}).

\bibitem[{\citenamefont{Kurucz et~al.}(2009)\citenamefont{Kurucz, S\o{}rensen,
  Taylor, Lukin, and Fleischhauer}}]{PhysRevLett.103.010502}
\bibinfo{author}{\bibfnamefont{Z.}~\bibnamefont{Kurucz}},
  \bibinfo{author}{\bibfnamefont{M.~W.} \bibnamefont{S\o{}rensen}},
  \bibinfo{author}{\bibfnamefont{J.~M.} \bibnamefont{Taylor}},
  \bibinfo{author}{\bibfnamefont{M.~D.} \bibnamefont{Lukin}}, \bibnamefont{and}
  \bibinfo{author}{\bibfnamefont{M.}~\bibnamefont{Fleischhauer}},
  \bibinfo{journal}{Phys. Rev. Lett.} \textbf{\bibinfo{volume}{103}},
  \bibinfo{pages}{010502} (\bibinfo{year}{2009}).

\bibitem[{\citenamefont{Witzel and Das~Sarma}(2007)}]{PhysRevB.76.045218}
\bibinfo{author}{\bibfnamefont{W.~M.} \bibnamefont{Witzel}} \bibnamefont{and}
  \bibinfo{author}{\bibfnamefont{S.}~\bibnamefont{Das~Sarma}},
  \bibinfo{journal}{Phys. Rev. B} \textbf{\bibinfo{volume}{76}},
  \bibinfo{pages}{045218} (\bibinfo{year}{2007}).

\bibitem[{\citenamefont{Wang and Zhang}(2012)}]{wang2012self}
\bibinfo{author}{\bibfnamefont{Q.}~\bibnamefont{Wang}} \bibnamefont{and}
  \bibinfo{author}{\bibfnamefont{Y.}~\bibnamefont{Zhang}},
  \bibinfo{journal}{Eur. Phys. J. B} \textbf{\bibinfo{volume}{85}},
  \bibinfo{pages}{225} (\bibinfo{year}{2012}).

\bibitem[{\citenamefont{Merkulov et~al.}(2002)\citenamefont{Merkulov, Efros,
  and Rosen}}]{Merkulov02}
\bibinfo{author}{\bibfnamefont{I.~A.} \bibnamefont{Merkulov}},
  \bibinfo{author}{\bibfnamefont{A.~L.} \bibnamefont{Efros}}, \bibnamefont{and}
  \bibinfo{author}{\bibfnamefont{M.}~\bibnamefont{Rosen}},
  \bibinfo{journal}{Phys. Rev. B} \textbf{\bibinfo{volume}{65}},
  \bibinfo{pages}{205309} (\bibinfo{year}{2002}).

\bibitem[{\citenamefont{Kane}(1998)}]{kane1998silicon}
\bibinfo{author}{\bibfnamefont{B.~E.} \bibnamefont{Kane}},
  \bibinfo{journal}{Nature (London)} \textbf{\bibinfo{volume}{393}},
  \bibinfo{pages}{133} (\bibinfo{year}{1998}).

\bibitem[{\citenamefont{Dobrovitski et~al.}(2006)\citenamefont{Dobrovitski,
  Taylor, and Lukin}}]{Dobrovitski06}
\bibinfo{author}{\bibfnamefont{V.~V.} \bibnamefont{Dobrovitski}},
  \bibinfo{author}{\bibfnamefont{J.~M.} \bibnamefont{Taylor}},
  \bibnamefont{and} \bibinfo{author}{\bibfnamefont{M.~D.} \bibnamefont{Lukin}},
  \bibinfo{journal}{Phys. Rev. B} \textbf{\bibinfo{volume}{73}},
  \bibinfo{pages}{245318} (\bibinfo{year}{2006}).

\bibitem[{\citenamefont{Salis et~al.}(2001)\citenamefont{Salis, Kato, Ensslin,
  Driscoll, Gossard, and Awschalom}}]{Salis2001}
\bibinfo{author}{\bibfnamefont{G.}~\bibnamefont{Salis}},
  \bibinfo{author}{\bibfnamefont{Y.}~\bibnamefont{Kato}},
  \bibinfo{author}{\bibfnamefont{K.}~\bibnamefont{Ensslin}},
  \bibinfo{author}{\bibfnamefont{D.~C.} \bibnamefont{Driscoll}},
  \bibinfo{author}{\bibfnamefont{A.~C.} \bibnamefont{Gossard}},
  \bibnamefont{and} \bibinfo{author}{\bibfnamefont{D.~D.}
  \bibnamefont{Awschalom}}, \bibinfo{journal}{Nature (London)}
  \textbf{\bibinfo{volume}{414}}, \bibinfo{pages}{619} (\bibinfo{year}{2001}).

\bibitem[{\citenamefont{Elzerman et~al.}(2004)\citenamefont{Elzerman, Hanson,
  Van~Beveren, Witkamp, Vandersypen, and Kouwenhoven}}]{elzerman2004single}
\bibinfo{author}{\bibfnamefont{J.}~\bibnamefont{Elzerman}},
  \bibinfo{author}{\bibfnamefont{R.}~\bibnamefont{Hanson}},
  \bibinfo{author}{\bibfnamefont{L.~W.} \bibnamefont{Van~Beveren}},
  \bibinfo{author}{\bibfnamefont{B.}~\bibnamefont{Witkamp}},
  \bibinfo{author}{\bibfnamefont{L.}~\bibnamefont{Vandersypen}},
  \bibnamefont{and} \bibinfo{author}{\bibfnamefont{L.~P.}
  \bibnamefont{Kouwenhoven}}, \bibinfo{journal}{Nature (London)}
  \textbf{\bibinfo{volume}{430}}, \bibinfo{pages}{431} (\bibinfo{year}{2004}).

\bibitem[{\citenamefont{Khaetskii et~al.}(2002)\citenamefont{Khaetskii, Loss,
  and Glazman}}]{Khaetskii2002}
\bibinfo{author}{\bibfnamefont{A.~V.} \bibnamefont{Khaetskii}},
  \bibinfo{author}{\bibfnamefont{D.}~\bibnamefont{Loss}}, \bibnamefont{and}
  \bibinfo{author}{\bibfnamefont{L.}~\bibnamefont{Glazman}},
  \bibinfo{journal}{Phys. Rev. Lett.} \textbf{\bibinfo{volume}{88}},
  \bibinfo{pages}{186802} (\bibinfo{year}{2002}).

\bibitem[{\citenamefont{Prokofev and Stamp}(2000)}]{Prokofev2000}
\bibinfo{author}{\bibfnamefont{N.~V.} \bibnamefont{Prokofev}} \bibnamefont{and}
  \bibinfo{author}{\bibfnamefont{P.}~\bibnamefont{Stamp}},
  \bibinfo{journal}{Rep. Prog. Phys} \textbf{\bibinfo{volume}{63}},
  \bibinfo{pages}{669} (\bibinfo{year}{2000}).

\bibitem[{\citenamefont{Gammon et~al.}(2001)\citenamefont{Gammon, Efros,
  Kennedy, Rosen, Katzer, Park, Brown, Korenev, and
  Merkulov}}]{gammon2001electron}
\bibinfo{author}{\bibfnamefont{D.}~\bibnamefont{Gammon}},
  \bibinfo{author}{\bibfnamefont{A.~L.} \bibnamefont{Efros}},
  \bibinfo{author}{\bibfnamefont{T.~A.} \bibnamefont{Kennedy}},
  \bibinfo{author}{\bibfnamefont{M.}~\bibnamefont{Rosen}},
  \bibinfo{author}{\bibfnamefont{D.~S.} \bibnamefont{Katzer}},
  \bibinfo{author}{\bibfnamefont{D.}~\bibnamefont{Park}},
  \bibinfo{author}{\bibfnamefont{S.~W.} \bibnamefont{Brown}},
  \bibinfo{author}{\bibfnamefont{V.~L.} \bibnamefont{Korenev}},
  \bibnamefont{and} \bibinfo{author}{\bibfnamefont{I.~A.}
  \bibnamefont{Merkulov}}, \bibinfo{journal}{Phys. Rev. Lett.}
  \textbf{\bibinfo{volume}{86}}, \bibinfo{pages}{5176} (\bibinfo{year}{2001}).

\bibitem[{\citenamefont{McCamey et~al.}(2009)\citenamefont{McCamey, van Tol,
  Morley, and Boehme}}]{dnp2009}
\bibinfo{author}{\bibfnamefont{D.~R.} \bibnamefont{McCamey}},
  \bibinfo{author}{\bibfnamefont{J.}~\bibnamefont{van Tol}},
  \bibinfo{author}{\bibfnamefont{G.~W.} \bibnamefont{Morley}},
  \bibnamefont{and} \bibinfo{author}{\bibfnamefont{C.}~\bibnamefont{Boehme}},
  \bibinfo{journal}{Phys. Rev. Lett.} \textbf{\bibinfo{volume}{102}},
  \bibinfo{pages}{027601} (\bibinfo{year}{2009}).

\bibitem[{\citenamefont{Asshoff et~al.}(2011)\citenamefont{Asshoff, W\"ust,
  Merz, Litvinov, Gerthsen, Kalt, and Hetterich}}]{Asshoff11}
\bibinfo{author}{\bibfnamefont{P.}~\bibnamefont{Asshoff}},
  \bibinfo{author}{\bibfnamefont{G.}~\bibnamefont{W\"ust}},
  \bibinfo{author}{\bibfnamefont{A.}~\bibnamefont{Merz}},
  \bibinfo{author}{\bibfnamefont{D.}~\bibnamefont{Litvinov}},
  \bibinfo{author}{\bibfnamefont{D.}~\bibnamefont{Gerthsen}},
  \bibinfo{author}{\bibfnamefont{H.}~\bibnamefont{Kalt}}, \bibnamefont{and}
  \bibinfo{author}{\bibfnamefont{M.}~\bibnamefont{Hetterich}},
  \bibinfo{journal}{Phys. Rev. B} \textbf{\bibinfo{volume}{84}},
  \bibinfo{pages}{125302} (\bibinfo{year}{2011}).

\bibitem[{\citenamefont{Paget et~al.}(1977)\citenamefont{Paget, Lampel,
  Sapoval, and Safarov}}]{PhysRevB.15.5780}
\bibinfo{author}{\bibfnamefont{D.}~\bibnamefont{Paget}},
  \bibinfo{author}{\bibfnamefont{G.}~\bibnamefont{Lampel}},
  \bibinfo{author}{\bibfnamefont{B.}~\bibnamefont{Sapoval}}, \bibnamefont{and}
  \bibinfo{author}{\bibfnamefont{V.~I.} \bibnamefont{Safarov}},
  \bibinfo{journal}{Phys. Rev. B} \textbf{\bibinfo{volume}{15}},
  \bibinfo{pages}{5780} (\bibinfo{year}{1977}).

\bibitem[{\citenamefont{Imamoglu
  et~al.}(2003)\citenamefont{Imamoglu, Knill,
  Tian, and Zoller}}]{imamoglu2003}
\bibinfo{author}{\bibfnamefont{A.}~\bibnamefont{Imamoglu}}, \bibinfo{author}{\bibfnamefont{E.}~\bibnamefont{Knill}},
  \bibinfo{author}{\bibfnamefont{L.}~\bibnamefont{Tian}}, \bibnamefont{and}
  \bibinfo{author}{\bibfnamefont{P.}~\bibnamefont{Zoller}},
  \bibinfo{journal}{Phys. Rev. Lett.} \textbf{\bibinfo{volume}{91}},
  \bibinfo{pages}{017402} (\bibinfo{year}{2003}).

\bibitem[{\citenamefont{Reilly et~al.}(2008)\citenamefont{Reilly, Taylor,
  Petta, Marcus, Hanson, and Gossard}}]{Reilly08}
\bibinfo{author}{\bibfnamefont{D.}~\bibnamefont{Reilly}},
  \bibinfo{author}{\bibfnamefont{J.}~\bibnamefont{Taylor}},
  \bibinfo{author}{\bibfnamefont{J.}~\bibnamefont{Petta}},
  \bibinfo{author}{\bibfnamefont{C.}~\bibnamefont{Marcus}},
  \bibinfo{author}{\bibfnamefont{M.}~\bibnamefont{Hanson}}, \bibnamefont{and}
  \bibinfo{author}{\bibfnamefont{A.}~\bibnamefont{Gossard}},
  \bibinfo{journal}{Science} \textbf{\bibinfo{volume}{321}},
  \bibinfo{pages}{817} (\bibinfo{year}{2008}).

\bibitem[{\citenamefont{Dixon et~al.}(1997)\citenamefont{Dixon, Wald, McEuen,
  and Melloch}}]{Dixon1997}
\bibinfo{author}{\bibfnamefont{D.~C.} \bibnamefont{Dixon}},
  \bibinfo{author}{\bibfnamefont{K.~R.} \bibnamefont{Wald}},
  \bibinfo{author}{\bibfnamefont{P.~L.} \bibnamefont{McEuen}},
  \bibnamefont{and} \bibinfo{author}{\bibfnamefont{M.~R.}
  \bibnamefont{Melloch}}, \bibinfo{journal}{Phys. Rev. B}
  \textbf{\bibinfo{volume}{56}}, \bibinfo{pages}{4743} (\bibinfo{year}{1997}).

\bibitem[{\citenamefont{Petta et~al.}(2008)\citenamefont{Petta, Taylor,
  Johnson, Yacoby, Lukin, Marcus, Hanson, and
  Gossard}}]{PhysRevLett.100.067601}
\bibinfo{author}{\bibfnamefont{J.~R.} \bibnamefont{Petta}},
  \bibinfo{author}{\bibfnamefont{J.~M.} \bibnamefont{Taylor}},
  \bibinfo{author}{\bibfnamefont{A.~C.} \bibnamefont{Johnson}},
  \bibinfo{author}{\bibfnamefont{A.}~\bibnamefont{Yacoby}},
  \bibinfo{author}{\bibfnamefont{M.~D.} \bibnamefont{Lukin}},
  \bibinfo{author}{\bibfnamefont{C.~M.} \bibnamefont{Marcus}},
  \bibinfo{author}{\bibfnamefont{M.~P.} \bibnamefont{Hanson}},
  \bibnamefont{and} \bibinfo{author}{\bibfnamefont{A.~C.}
  \bibnamefont{Gossard}}, \bibinfo{journal}{Phys. Rev. Lett.}
  \textbf{\bibinfo{volume}{100}}, \bibinfo{pages}{067601}
  (\bibinfo{year}{2008}).

\bibitem[{\citenamefont{Gullans et~al.}(2010)\citenamefont{Gullans, Krich,
  Taylor, Bluhm, Halperin, Marcus, Stopa, Yacoby, and Lukin}}]{Lukin10}
\bibinfo{author}{\bibfnamefont{M.}~\bibnamefont{Gullans}},
  \bibinfo{author}{\bibfnamefont{J.~J.} \bibnamefont{Krich}},
  \bibinfo{author}{\bibfnamefont{J.~M.} \bibnamefont{Taylor}},
  \bibinfo{author}{\bibfnamefont{H.}~\bibnamefont{Bluhm}},
  \bibinfo{author}{\bibfnamefont{B.~I.} \bibnamefont{Halperin}},
  \bibinfo{author}{\bibfnamefont{C.~M.} \bibnamefont{Marcus}},
  \bibinfo{author}{\bibfnamefont{M.}~\bibnamefont{Stopa}},
  \bibinfo{author}{\bibfnamefont{A.}~\bibnamefont{Yacoby}}, \bibnamefont{and}
  \bibinfo{author}{\bibfnamefont{M.~D.} \bibnamefont{Lukin}},
  \bibinfo{journal}{Phys. Rev. Lett.} \textbf{\bibinfo{volume}{104}},
  \bibinfo{pages}{226807} (\bibinfo{year}{2010}).

\bibitem[{\citenamefont{Zhang et~al.}(2010)\citenamefont{Zhang, Hu, Zhuang,
  You, and Liu}}]{Zhang10}
\bibinfo{author}{\bibfnamefont{W.}~\bibnamefont{Zhang}},
  \bibinfo{author}{\bibfnamefont{J.-L.} \bibnamefont{Hu}},
  \bibinfo{author}{\bibfnamefont{J.}~\bibnamefont{Zhuang}},
  \bibinfo{author}{\bibfnamefont{J.~Q.} \bibnamefont{You}}, \bibnamefont{and}
  \bibinfo{author}{\bibfnamefont{R.-B.} \bibnamefont{Liu}},
  \bibinfo{journal}{Phys. Rev. B} \textbf{\bibinfo{volume}{82}},
  \bibinfo{pages}{045314} (\bibinfo{year}{2010}).

\bibitem[{\citenamefont{Wu et~al.}(2013)\citenamefont{Wu, Ding, Shi, and
  Zhang}}]{Zhang13arxiv}
\bibinfo{author}{\bibfnamefont{N.}~\bibnamefont{Wu}},
  \bibinfo{author}{\bibfnamefont{W.}~\bibnamefont{Ding}},
  \bibinfo{author}{\bibfnamefont{A.}~\bibnamefont{Shi}}, \bibnamefont{and}
  \bibinfo{author}{\bibfnamefont{W.}~\bibnamefont{Zhang}},
  \bibinfo{journal}{arXiv:1303.0590v1 [cond-mat.mes-hall]}
  (\bibinfo{year}{2013}).

\bibitem[{\citenamefont{Nielsen and Chuang}(2010)}]{nielsen2010quantum}
\bibinfo{author}{\bibfnamefont{M.~A.} \bibnamefont{Nielsen}} \bibnamefont{and}
  \bibinfo{author}{\bibfnamefont{I.~L.} \bibnamefont{Chuang}},
  \emph{\bibinfo{title}{Quantum computation and quantum information}}
  (\bibinfo{publisher}{Cambridge university press}, \bibinfo{year}{2010}).

\bibitem[{\citenamefont{Bensky et~al.}(2012)\citenamefont{Bensky, Petrosyan,
  Majer, Schmiedmayer, and Kurizki}}]{opinhomo12}
\bibinfo{author}{\bibfnamefont{G.}~\bibnamefont{Bensky}},
  \bibinfo{author}{\bibfnamefont{D.}~\bibnamefont{Petrosyan}},
  \bibinfo{author}{\bibfnamefont{J.}~\bibnamefont{Majer}},
  \bibinfo{author}{\bibfnamefont{J.}~\bibnamefont{Schmiedmayer}},
  \bibnamefont{and} \bibinfo{author}{\bibfnamefont{G.}~\bibnamefont{Kurizki}},
  \bibinfo{journal}{Phys. Rev. A} \textbf{\bibinfo{volume}{86}},
  \bibinfo{pages}{012310} (\bibinfo{year}{2012}).

\bibitem[{\citenamefont{Diniz et~al.}(2011)\citenamefont{Diniz, Portolan,
  Ferreira, G\'erard, Bertet, and Auff\`eves}}]{PhysRevA.84.063810}
\bibinfo{author}{\bibfnamefont{I.}~\bibnamefont{Diniz}},
  \bibinfo{author}{\bibfnamefont{S.}~\bibnamefont{Portolan}},
  \bibinfo{author}{\bibfnamefont{R.}~\bibnamefont{Ferreira}},
  \bibinfo{author}{\bibfnamefont{J.~M.} \bibnamefont{G\'erard}},
  \bibinfo{author}{\bibfnamefont{P.}~\bibnamefont{Bertet}}, \bibnamefont{and}
  \bibinfo{author}{\bibfnamefont{A.}~\bibnamefont{Auff\`eves}},
  \bibinfo{journal}{Phys. Rev. A} \textbf{\bibinfo{volume}{84}},
  \bibinfo{pages}{063810} (\bibinfo{year}{2011}).

\bibitem[{\citenamefont{Johnson et~al.}(2005)\citenamefont{Johnson, Petta,
  Taylor, Yacoby, Lukin, Marcus, Hanson, and Gossard}}]{johnson2005triplet}
\bibinfo{author}{\bibfnamefont{A.}~\bibnamefont{Johnson}},
  \bibinfo{author}{\bibfnamefont{J.}~\bibnamefont{Petta}},
  \bibinfo{author}{\bibfnamefont{J.}~\bibnamefont{Taylor}},
  \bibinfo{author}{\bibfnamefont{A.}~\bibnamefont{Yacoby}},
  \bibinfo{author}{\bibfnamefont{M.}~\bibnamefont{Lukin}},
  \bibinfo{author}{\bibfnamefont{C.}~\bibnamefont{Marcus}},
  \bibinfo{author}{\bibfnamefont{M.}~\bibnamefont{Hanson}}, \bibnamefont{and}
  \bibinfo{author}{\bibfnamefont{A.}~\bibnamefont{Gossard}},
  \bibinfo{journal}{Nature (London)} \textbf{\bibinfo{volume}{435}},
  \bibinfo{pages}{925} (\bibinfo{year}{2005}).

\bibitem[{\citenamefont{Koppens et~al.}(2005)\citenamefont{Koppens, Folk,
  Elzerman, Hanson, van Beveren, Vink, Tranitz, Wegscheider, Kouwenhoven, and
  Vandersypen}}]{Koppens2005}
\bibinfo{author}{\bibfnamefont{F.}~\bibnamefont{Koppens}},
  \bibinfo{author}{\bibfnamefont{J.}~\bibnamefont{Folk}},
  \bibinfo{author}{\bibfnamefont{J.}~\bibnamefont{Elzerman}},
  \bibinfo{author}{\bibfnamefont{R.}~\bibnamefont{Hanson}},
  \bibinfo{author}{\bibfnamefont{L.}~\bibnamefont{van Beveren}},
  \bibinfo{author}{\bibfnamefont{I.}~\bibnamefont{Vink}},
  \bibinfo{author}{\bibfnamefont{H.}~\bibnamefont{Tranitz}},
  \bibinfo{author}{\bibfnamefont{W.}~\bibnamefont{Wegscheider}},
  \bibinfo{author}{\bibfnamefont{L.}~\bibnamefont{Kouwenhoven}},
  \bibnamefont{and}
  \bibinfo{author}{\bibfnamefont{L.}~\bibnamefont{Vandersypen}},
  \bibinfo{journal}{Science} \textbf{\bibinfo{volume}{309}},
  \bibinfo{pages}{1346} (\bibinfo{year}{2005}).

\bibitem[{\citenamefont{Petta et~al.}(2005)\citenamefont{Petta, Johnson,
  Taylor, Laird, Yacoby, Lukin, Marcus, Hanson, and Gossard}}]{Petta2005}
\bibinfo{author}{\bibfnamefont{J.}~\bibnamefont{Petta}},
  \bibinfo{author}{\bibfnamefont{A.}~\bibnamefont{Johnson}},
  \bibinfo{author}{\bibfnamefont{J.}~\bibnamefont{Taylor}},
  \bibinfo{author}{\bibfnamefont{E.}~\bibnamefont{Laird}},
  \bibinfo{author}{\bibfnamefont{A.}~\bibnamefont{Yacoby}},
  \bibinfo{author}{\bibfnamefont{M.}~\bibnamefont{Lukin}},
  \bibinfo{author}{\bibfnamefont{C.}~\bibnamefont{Marcus}},
  \bibinfo{author}{\bibfnamefont{M.}~\bibnamefont{Hanson}}, \bibnamefont{and}
  \bibinfo{author}{\bibfnamefont{A.}~\bibnamefont{Gossard}},
  \bibinfo{journal}{Science} \textbf{\bibinfo{volume}{309}},
  \bibinfo{pages}{2180} (\bibinfo{year}{2005}).

\bibitem[{\citenamefont{Gullans et~al.}(2013)\citenamefont{Gullans, Krich,
  Taylor, Halperin, and Lukin}}]{Lukin13}
\bibinfo{author}{\bibfnamefont{M.}~\bibnamefont{Gullans}},
  \bibinfo{author}{\bibfnamefont{J.~J.} \bibnamefont{Krich}},
  \bibinfo{author}{\bibfnamefont{J.~M.} \bibnamefont{Taylor}},
  \bibinfo{author}{\bibfnamefont{B.~I.} \bibnamefont{Halperin}},
  \bibnamefont{and} \bibinfo{author}{\bibfnamefont{M.~D.} \bibnamefont{Lukin}},
  \bibinfo{journal}{Phys. Rev. B} \textbf{\bibinfo{volume}{88}},
  \bibinfo{pages}{035309} (\bibinfo{year}{2013}).

\bibitem[{\citenamefont{Yao and Luo}(2010)}]{Yao_EPL_2010}
\bibinfo{author}{\bibfnamefont{W.}~\bibnamefont{Yao}} \bibnamefont{and}
  \bibinfo{author}{\bibfnamefont{Y.}~\bibnamefont{Luo}}, \bibinfo{journal}{EPL}
  \textbf{\bibinfo{volume}{92}}, \bibinfo{pages}{17008} (\bibinfo{year}{2010}).

\bibitem[{\citenamefont{Coish and Loss}(2004)}]{PhysRevB.70.195340}
\bibinfo{author}{\bibfnamefont{W.~A.} \bibnamefont{Coish}} \bibnamefont{and}
  \bibinfo{author}{\bibfnamefont{D.}~\bibnamefont{Loss}},
  \bibinfo{journal}{Phys. Rev. B} \textbf{\bibinfo{volume}{70}},
  \bibinfo{pages}{195340} (\bibinfo{year}{2004}).

\bibitem[{\citenamefont{Bortz and Stolze}(2007)}]{exsolution07}
\bibinfo{author}{\bibfnamefont{M.}~\bibnamefont{Bortz}} \bibnamefont{and}
  \bibinfo{author}{\bibfnamefont{J.}~\bibnamefont{Stolze}},
  \bibinfo{journal}{Phys. Rev. B} \textbf{\bibinfo{volume}{76}},
  \bibinfo{pages}{014304} (\bibinfo{year}{2007}).

\bibitem[{\citenamefont{Gaudin}(1976)}]{gaudin1976diagonalization}
\bibinfo{author}{\bibfnamefont{M.}~\bibnamefont{Gaudin}}, \bibinfo{journal}{J.
  Phys. (Paris)} \textbf{\bibinfo{volume}{37}}, \bibinfo{pages}{1087}
  (\bibinfo{year}{1976}).

\bibitem [{\citenamefont {Faribault}\ and\ \citenamefont
  {Schuricht}(2013)}]{Faribault13}
  \bibfield  {author} {\bibinfo {author} {\bibfnamefont {A.}~\bibnamefont
  {Faribault}}\ and\ \bibinfo {author} {\bibfnamefont {D.}~\bibnamefont
  {Schuricht}},\ }{\bibfield  {journal} {\bibinfo  {journal}
  {Phys. Rev. Lett.}\ }\textbf {\bibinfo {volume} {110}},\ \bibinfo {pages}
  {040405} (\bibinfo {year} {2013})}.

\bibitem[{\citenamefont{Dobrovitski and De~Raedt}(2003)}]{Dobrovitski03}
\bibinfo{author}{\bibfnamefont{V.~V.} \bibnamefont{Dobrovitski}}
  \bibnamefont{and} \bibinfo{author}{\bibfnamefont{H.~A.}
  \bibnamefont{De~Raedt}}, \bibinfo{journal}{Phys. Rev. E}
  \textbf{\bibinfo{volume}{67}}, \bibinfo{pages}{056702}
  (\bibinfo{year}{2003}).

\end{thebibliography}

\end{document}